\documentclass[conference]{IEEEtran}
\IEEEoverridecommandlockouts
\usepackage{subcaption}
\usepackage{cite}
\usepackage{amsmath,amssymb,amsfonts}
\usepackage{algorithmic}
\usepackage{graphicx}
\usepackage{textcomp}
\usepackage{xcolor}
\usepackage{float}
\usepackage{hyperref}
\usepackage{rotating} 
\usepackage{subcaption} 
\usepackage{caption}  
\usepackage{xcolor}   

\usepackage{array, multirow}
\usepackage{xurl}

\def\BibTeX{{\rm B\kern-.05em{\sc i\kern-.025em b}\kern-.08em
    T\kern-.1667em\lower.7ex\hbox{E}\kern-.125emX}}

\begin{document}

\title{Precision Enhancement in Sustained Visual Attention Training Platforms: Offline EEG Signal Analysis for Classifier Fine-Tuning\\

\thanks{This project was supported by the Rhode Island INBRE program from the National Institute of General Medical Sciences of the National Institutes of Health under grant number P20GM103430 and National Science Foundation under award ID 2245558. Author 4 is the corresponding author.
}
}



\author{\IEEEauthorblockN{1\textsuperscript{st} Maryam Norouzi}
\IEEEauthorblockA{\textit{Dept. of Electrical, Computer and Biomedical Engineering} \\
\textit{University of Rhode Island}\\
Kingston, RI, USA \\
maryam\_norouzi@uri.edu}
\and
\IEEEauthorblockN{2\textsuperscript{nd} Mohammad Zaeri Amirani}
\IEEEauthorblockA{\textit{Dept. of Electrical, Computer and Biomedical Engineering} \\
\textit{University of Rhode Island}\\
Kingston, RI, USA \\
zaeri.amirani@uri.edu}
\and
\IEEEauthorblockN{3\textsuperscript{rd} Yalda Shahriari}
\IEEEauthorblockA{\textit{Dept. of Electrical, Computer and Biomedical Engineering} \\
\textit{University of Rhode Island}\\
Kingston, RI, USA \\
yalda\_shahriari@uri.edu\\
~\IEEEmembership{~member, IEEE}}
\and
\IEEEauthorblockN{4\textsuperscript{th} Reza Abiri}
\IEEEauthorblockA{\textit{Dept. of Electrical, Computer and Biomedical Engineering} \\
\textit{University of Rhode Island}\\
Kingston, RI, USA \\
reza\_abiri@uri.edu\\
~\IEEEmembership{~member, IEEE}}
}

\maketitle
\begin{abstract}
In this study, a novel open-source brain-computer interface (BCI) platform was developed to decode scalp electroencephalography (EEG) signals associated with sustained attention. The EEG signal collection was conducted using a wireless headset during a sustained visual attention task, where participants were instructed to discriminate between composite images superimposed with scenes and faces, responding only to the relevant subcategory while ignoring the irrelevant ones. Seven volunteers participated in  this experiment. The data collected were subjected to analyses through event-related potential (ERP), Hilbert Transform, and Wavelet Transform to extract temporal and spectral features. For each participant, utilizing its extracted features, personalized Support Vector Machine (SVM) and Random Forest (RF) models with tuned hyperparameters were developed. The models aimed to decode the participant's attentional state towards the face and scene stimuli. The SVM models achieved a higher average accuracy of 80\% and an Area Under the Curve (AUC) of 0.86, while the RF models showed an average accuracy of 78\% and AUC of 0.8. This work suggests potential applications for the evaluation of visual attention and the development of closed-loop brainwave regulation systems in the future.

\end{abstract} 

\IEEEpeerreviewmaketitle

\section{Introduction}
Attention is a fundamental aspect of human cognition and perception \cite{yoo2022brain}. Traditionally, attentional states have been studied using blood-oxygen-level dependent signals via functional magnetic resonance imaging \cite{thibault2016self}, which, despite its high spatial resolution, has limitations for real-time neurofeedback due to the slow vascular response of the brain \cite{nicolas2012brain}.In contrast, electroencephalography (EEG), due to its higher temporal resolution is more suitable for real-time brain-computer interface (BCI) applications \cite{abiri2019comprehensive}. EEG has been widely used for attention evaluation and training, particularly in children with attention deficit hyperactivity disorders \cite{lim2023home}. Since face-like visual stimuli undergo specialized processing in the human brain \cite{rossion2023intracerebral}, face images were employed in numerous brain studies, particularly studies on attention \cite{tuckute2021real}. Among them, the work of Sreenivasan et al. is notable \cite{sreenivasan2009attention}. They employed event-related potential (ERP) analysis to show that attention to faces in composite images with different transparency of face can modulate the perceptual processing of faces. The other recent study explored the spatiotemporal changes in EEG to classify perceptual states (faces/Gabors) and analyzed the scope of attention (locally/globally) \cite{list2017pattern}; nevertheless, it did not explored the EEG classification results for when subjects focused on both image categories within a sequence of overlapped images. Based on our investigation, most previous works focused on identifying participants' attention levels without examining the visual stimuli that trigger them. Recently, Abiri et al. focused on decoding EEG-based attentional states towards faces and scenes. They developed a BCI platform using MATLAB, incorporating a linear SVM classifier. However, this classifier did not include subject-specific parameter tuning, which is particularly crucial for real-time neurofeedback platforms\cite{abiri2019comprehensive}. To the best of the authors' knowledge, no studies have yet reported on the development and hyperparameter tuning of non-linear ML classifiers to assess a participant's sustained visual attention and to identify the triggering visual stimulus. To address this, in this study, we introduced an innovative wireless, portable, and open-source EEG-based BCI software\footnote{\url{https://github.com/AbiriLab/Neurofeedback-Based-BCI}}, to explore subjects' sustained visual attention levels toward two stimulus categories—faces and scenes. We employed ERP, Morlet continuous Wavelet Transform (MCWT), and Hilbert Transform (HT) analyses to discern distinctive neural patterns and extract relevant features. Subsequently, to assess the level of sustained visual attention in a participant, along with identifying the specific visual stimulus responsible for capturing that attention, two classifiers were developed, fine-tuned, and personalized using the extracted features as input.
\section{Materials and methods}
\subsection{EEG-Based BCI Platform: Recording Device and Interface}
\begin{figure*}[ht]
  \centering
  \begin{minipage}{.5\textwidth}
    \centering
    \includegraphics[width=1\textwidth, trim={.7cm 1.4cm 1cm 3.5cm}, clip]{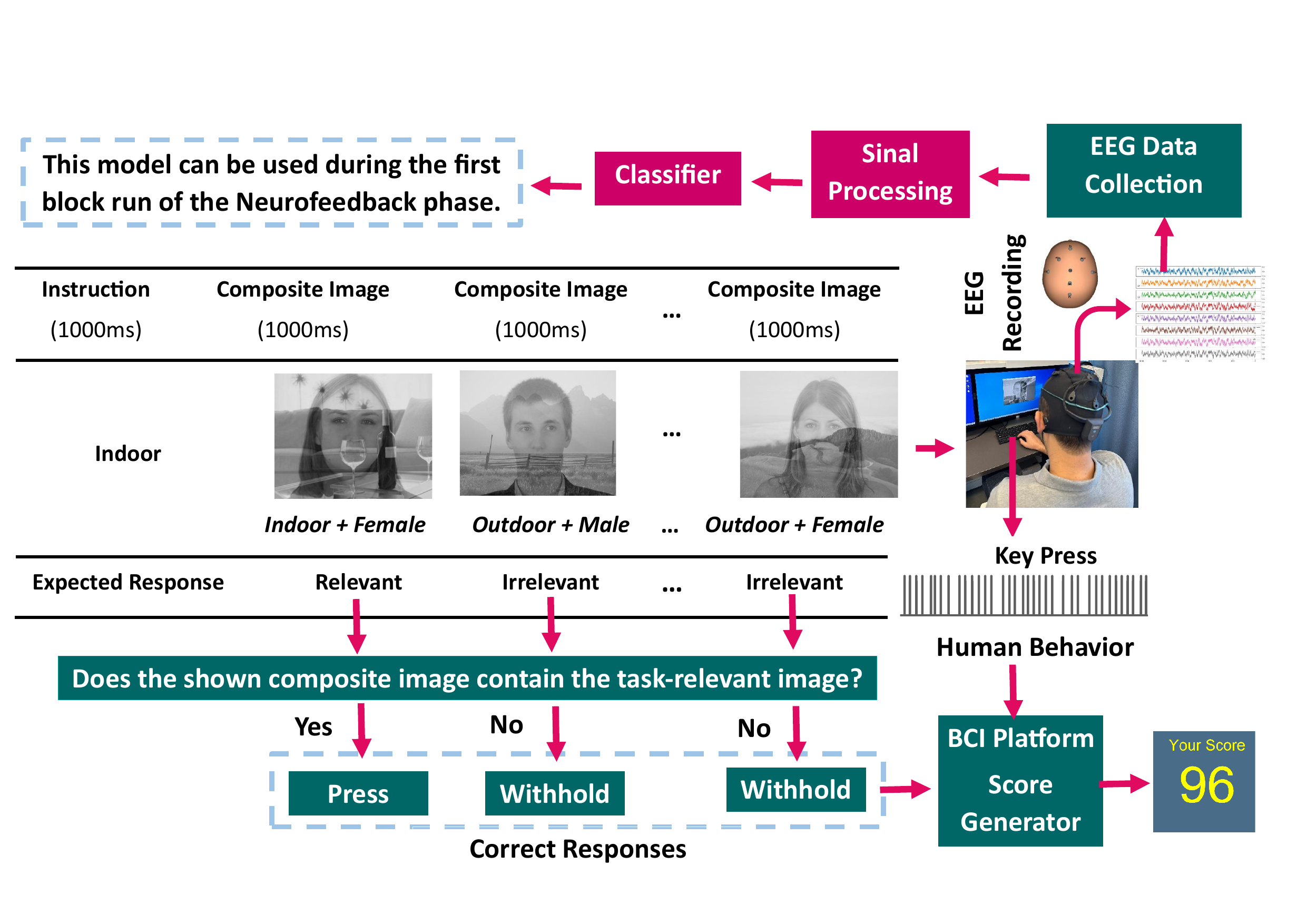}
  \end{minipage}%
  \begin{minipage}{.5\textwidth}
    \centering
    \includegraphics[width=.85\textwidth, trim={1cm 1.5cm 1cm .4cm}, clip]{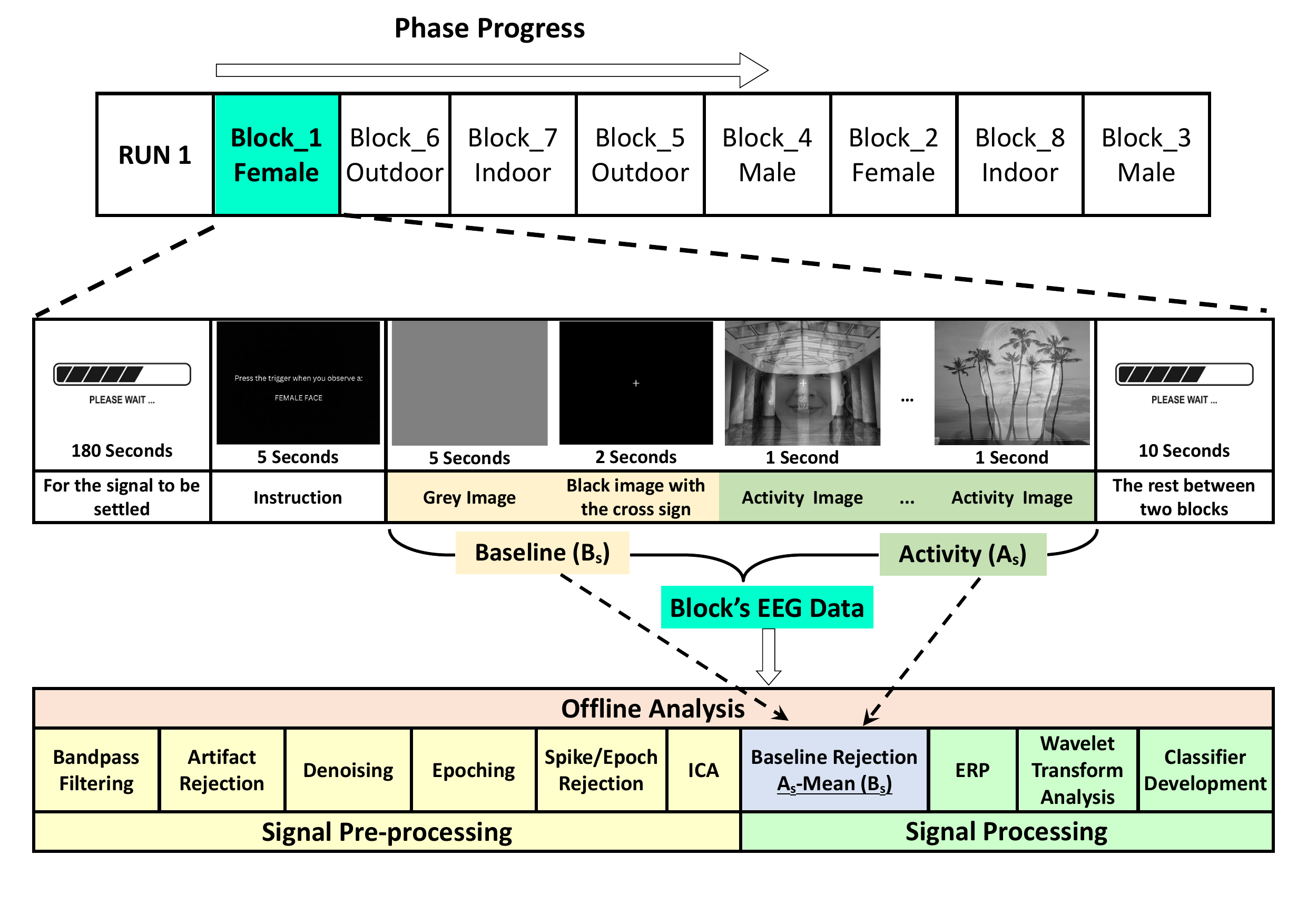}
  \end{minipage}
  \caption{ Developed BCI Platform (Clinical Trial: NCT05908253). Left: A sample sequence of composite images during one block and the corresponding expected responses. Right: Flowchart of the experiment and data analysis}
  \label{fig:bci-platform}
\end{figure*}

The BCI platform includes a wireless EEG headset, a workstation equipped with dual monitors (experimenter and participant screens for control and stimuli), and a novel developed Python-based software for simultaneous data acquisition, analysis, and execution of a designed paradigm (Fig. \ref{fig:bci-platform}). EEG signals were acquired at 250 Hz using Unicorn Hybrid Black (g.tec medical engineering GmbH), and transmitted to the PC via Bluetooth.The headset has 8 electrodes at Fz, C3, Cz, C4, Pz, Po7, Oz, Po8 (10-20 system), plus two snap electrodes for reference and ground behind the left and right ear, respectively \cite{unicorn_bi}. 

\subsection{Experimental Protocol and Subjects}
Seven healthy participants (4 males and 3 females) with a mean age of 28.14 years and a standard deviation of 10.84 took part in this study. All participants had normal or corrected-to-normal vision. After the Institutional Review Board (IRB) approval process, all participants gave written informed consent to the experiment. The participants were asked to sit comfortably in a fixed chair with one hand resting on their lap and another hand ready to press a key to give behavioral responses. The participants were instructed to pay attention to the images displayed on the monitor during the experiment and limit their body movement. In the present work, we aimed to identify participants’ attentional states toward two categories of images: face versus scene; regardless of their subcategories. We assumed that neural response contains common features for the subcategories within a single category \cite{abiri2019decoding}. 

In our study, we employed the Sustained Attention to Response Task (SART) paradigm \cite{robertson1997oops}. Our paradigm consists of eight blocks of trials with a 10-second respite between blocks. The experiment starts with a 3-minute EEG recording for signal stabilization, followed by the blocks of trials in a random order.(Fig. \ref{fig:bci-platform}).  Each block begins with a five-second texture cue, then a 7-second grey image (baseline), followed by forty 1-second image stimuli trials (activity trials). Considering all eight blocks, this paradigm results in a total of 320 activity trials throughout the experiment. In our experiment, four subcategories of indoor and outdoor scenes and male and female face images were chosen as stimuli. The images were all black and white with equal sizes, 800 × 1000 pixels. Face images were chosen to be neutral and were centered inside the composite image. The indoor images were chosen from interior scenes. Outdoor images were natural landscapes and cityscapes. Brightness and contrast for all face and scene images were adjusted so that the images have equalized contrast. Each trial includes a composite image with equal proportions (50\%) of the scene and face. (In our future neurofeedback task, the proportions of images in the composite images will be regulated). There was no repetition of face or scene images through each block of the experiment, which helps to prevent any learning mechanism for the participant \cite{hannula2012hippocampus}. Participants were asked to identify whether the shown image contained a task-relevant/irrelevant image (e.g., an indoor/outdoor image). They were asked to press the key on the keyboard for each recognized relevant image and withhold their responses for irrelevant images. Following the SART paradigm, which involves the withholding of key presses to rare targets, 90\% of the composite images contained images from the task-relevant subcategory. Each participant underwent a total of approximately 20 minutes of experimentation, including EEG cap fitting and electrode impedance checks.

\subsection{Signal Pre-processing }
 To do pre-processing, each EEG channel underwent a series of preprocessing steps. Firstly, A 5th-order Butterworth bandpass filter isolated 0.4–40 Hz frequencies. Subsequently, spikes were detected using the Median Absolute Deviation technique and interpolated via a cubic spline, maintaining EEG data integrity. Then, to minimize random noise a K-neighbors regression was used. Next, baseline correction was implemented by subtracting the average of baseline signals in each block from the activity signals of that block to enhance task-specific brain response clarity. Finally, data was normalized across experiment blocks using Z-scoring.

\subsection{Signal Processing, Feature Extraction, and Classification}
Following our initial analysis, we identified several ERPs linked to sustained visual attention. Additionally, the Hilbert Envelope (HE) revealed distinct patterns for face and scene during sustained visual attention tasks. The HE highlights temporal changes in EEG signal amplitude but not detailed frequency-based oscillations. In contrast, the MCWT  excels at extracting transient features from non-stationary data, making it valuable for EEG data analysis\cite{deng2012massively}. Combining MCWT with the HE offers a thorough approach essential for robust attentional state modeling. So, in this study, ERP, Hilbert envelop, and the MCWT were employed for the analysis of EEG data. For our study's binary classification, SVM was selected due to its proven effectiveness in face recognition tasks \cite{shtino2023comparative}. Additionally, acknowledging Random Forest's (RF) capabilities in processing non-linear data, resisting overfitting, accommodating missing values, and its scalability, efficiency, and proficiency with imbalanced datasets, we developed an RF model to assess and compare its classification performance with SVM \cite{siemers2023differences}. The parameters of both classifiers also have been tuned.

\subsubsection{Feature Extraction}
\begin{figure*}[t]
  \centering
  \begin{minipage}{.27\textwidth}
    \centering
    \includegraphics[width=1.2\linewidth]{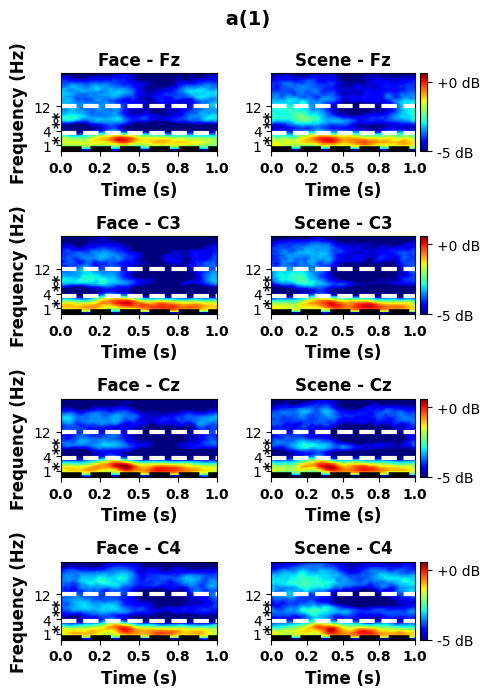}
  \end{minipage}%
  \begin{minipage}{.23\textwidth}
    \centering
    \includegraphics[width=.85\linewidth]{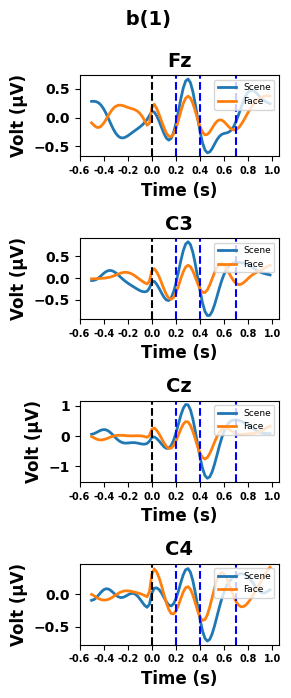}
  \end{minipage}%
  \begin{minipage}{.27\textwidth}
    \centering
    \includegraphics[width=1.2\linewidth]{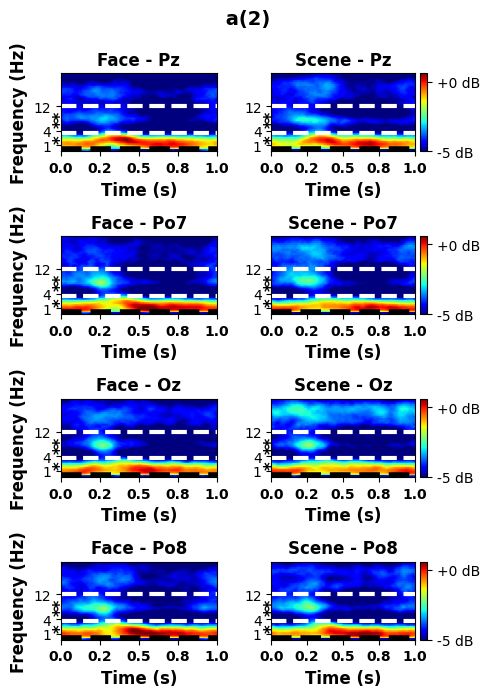} 
  \end{minipage}%
  \begin{minipage}{.23\textwidth}
    \centering
    \includegraphics[width=.85\linewidth]{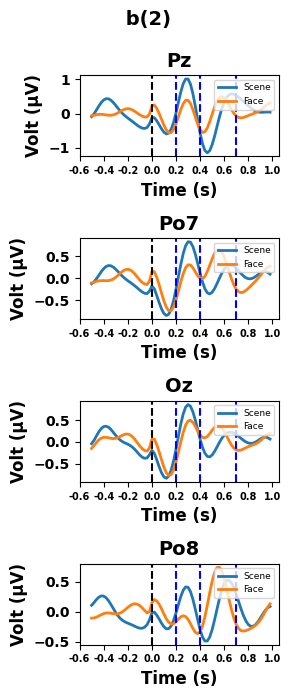} 
  \end{minipage}
  \caption{a(1) and a(2): Time-frequency power spectra and b(1) and b(2) ERP analyses results by channel for one subject, "*" delta band, "**" theta and early alpha band }
  \label{fig:test}
\end{figure*}

 ERP features were extracted through a series of steps: bandpass filtering (1-4 Hz), segmentation into epochs (each block was segmented into 250-sample epochs, each corresponding to 1000 milliseconds), and downsampling to 50 Hz (50 samples per epoch to improve the signal-to-noise ratio). The epochs were then averaged to derive distinct mean signals for baseline and stimulus-evoked activities per block. Next, the averages for each category were calculated to identify the distinct ERP waveforms for face and scene stimuli (Fig. \ref{fig:test}(a)). For extracting ERP-related features, key statistical parameters such as mean amplitude, variance, standard deviation, peak-to-peak amplitude, zero crossings, and number of peaks were computed within the identified time windows: early response (0-50 ms), mid-range processing (80-210 ms, 240-350 ms, 400-500 ms, 520-630 ms), and extended analysis periods (650-900 ms), with an overall assessment across the full epoch (0-1000 ms) leading to 42 (6x7) ERP features for each channel.  Also, utilizing Linear Discriminant Analysis, the intra-class and inter-class variance have been minimized and maximized, respectively, yielding a single, feature per EEG channel. Overall, the analysis led to the extraction of a total of 344 (8x43) ERP-related features for each trial.
To extract time-frequency features, MCWT were generated for 0-40 Hz frequencies in 1 Hz increments, with cycles linearly varying between 0.1 and 10, and convolved with EEG data. Next, epoching and baseline normalization were performed, quantified using the logarithmic equation \(dB = 10 \cdot \log_{10} \left( \frac{\text{activity}}{\text{baseline}} \right)\), where \(dB\) represents the power in decibels. Finally, for each epoch of activity, signal mean, variance, peak frequency, peak magnitude, skewness, energy, and kurtosis, were extracted which led to 56 (8x7) time-frequency features for each trial.
In addition, the Hilbert transform was applied to extract the envelope of EEG data in various frequency bands (Delta: [1, 4] Hz, theta: [4, 8] Hz, alpha: [8, 14] Hz, beta: [14, 30] Hz,  and gamma: [30, 40] Hz). Then the mean, median, standard deviation, skewness, energy, and kurtosis of each envelope computed led to the extraction of 240 (8x30) features per trial.  

\subsubsection{Classifier Development}
To develop classifiers, initially, all features were combined to form a comprehensive feature matrix containing 640 features per trial across a total of 320 trials. Then classifiers were developed and their hyperparameters systemically optimized for each subject using the well-established Optuna library \cite{akiba2019optuna}, ensuring that the selections were not random but systemically determined. Optuna employs Bayesian optimization with a Tree-structured Parzen Estimator (TPE) to efficiently explore the hyperparameter space. For each subject in our study, a separate Optuna experiment was created, and the hyperparameters were optimized individually. This approach ensures that the model is finely tuned to each subject's unique EEG data characteristics, maximizing the predictive accuracy and generalization capability of the model. The criteria for selecting the range of hyperparameters were based on preliminary empirical tests and literature review, ensuring they are suitable for the complexity and nature of EEG data being analyzed. We aimed to define a comprehensive yet computationally feasible hyperparameter range with the TPE algorithm, maximizing cross-validation accuracy and minimizing overfitting to ensure robust model performance on unseen data. When tuning the RF classifier, the hyperparameters that were used are as follows: (i) the number of estimators was set between 2 and 10, (ii) the depth of the tree was between 5 and 20, (iii) the minimum number of samples required to split an internal node was between 2 and 20, (iv) minimum number of samples to qualify as a leaf node was between 2 and 5. Moreover, when looking for the best split, the number of features considered from [‘auto’, ‘sqrt’, ‘log 2’] and [‘gini’, ‘entropy’] was used to measure the quality of the split (criterion). Hyperparameters for SVM tuning—misclassification cost ($C$) and Gaussian kernel parameter ($\Gamma$)—were bounded by a predefined logarithmic scale ranging from 0.001 to 1000, to balance the decision boundary and the complexity of the kernel. Validation of the final model was done based on the Receiver Operating Characteristic (ROC) curves, AUC scores, and mean accuracy metrics (Table \ref{tab:optimization_results} and Fig. \ref{fig:auc}).

\section{Results and discussion}
\subsection{Time and Frequency Domain Analysis}
The ERP waveforms (Fig. \ref{fig:test} (b1) and (b2)), reveal distinct temporal patterns between scene and face stimuli across EEG channels (early, mid-range, and extended
response). As it is shown in the wavelet power spectra (Fig. \ref{fig:test} (a1) and (a2)), there is a prominent power concentration in the lower frequency bands (Delta and Theta and early Alpha) during the early to mid-time windows (0-800 ms), particularly for face stimuli. In contrast to this, scene processing appears more distributed across frequencies and time. The HE also indicates distinct patterns in brain activity corresponding to the face and scene conditions, particularly in the Alpha and Beta bands.




\begin{table}[tp]
\centering
\caption{Models' tuned hyperparameter and performance across subjects. Abbreviations: NS: Number of Estimators, MD: Max Depth, MSS: Min Samples Split, MF: Max Features.}
\label{tab:optimization_results}
\footnotesize 
\begin{tabular}{>{\centering\arraybackslash}m{.5cm}>{\centering\arraybackslash}m{.9cm}p{0.4cm}p{0.4cm}p{0.4cm}p{0.4cm}p{0.4cm}p{0.4cm}p{0.8cm}}
\hline
Model & Param. & \multicolumn{7}{c}{Subjects}  \\ 
\cline{3-9}
 & & 1 & 2 & 3 & 4 & 5 & 6 & 7\\ 
\hline
\multirow{5}{*}{RF} & NS & 9 & 9 & 10 & 8 & 10 & 8 & 8  \\
 & MD & 13 & 5 & 7 & 11 & 5 & 10 & 5  \\
 & MSS & 13 & 6 & 8 & 16 & 19 & 7 & 9  \\
 & MF & auto & sqrt & sqrt & sqrt & sqrt & auto & auto  \\
 & ACC & 75\% & 79\% & 72\% & 80\% & 81\% & 80\% & \textbf{78}\% \\
 & AUC & 0.72 & 0.83 & 0.70 & 0.82 & 0.84 & 0.87 & 0.84\\ 
\hline
\multirow{4}{*}{SVM} & $C$ & 93 & 354 & 672 & 188 & 894 & 20 & 1.5e-2\\
& $\Gamma$$e-3$ & 1.8 & 1 & 2 & 1 & 1.1 & 1 & 1\\
 & ACC & \textbf{79}\% & \textbf{82}\% & \textbf{73}\% & 80\% & \textbf{85}\% & \textbf{84}\% & 75\% \\
 & AUC & \textbf{0.84} & \textbf{0.89} & \textbf{0.76} & \textbf{0.85} & \textbf{0.90} & \textbf{0.92} & \textbf{0.88} \\ 
\hline
\end{tabular}
\end{table}


\begin{figure}[h]
  \centering
    \centering
    \includegraphics[width=.4\textwidth, clip]{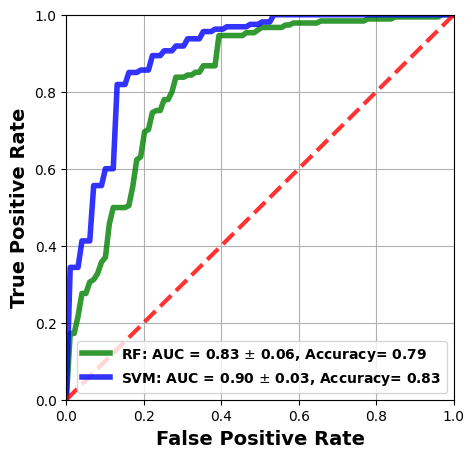}
  \caption{ROC curves of the tuned Scene and Face Classifiers}
  \label{fig:auc}
\end{figure}

\subsection{Personalized classifier tuning}
Table \ref{tab:optimization_results} summarizes the results of the hyperparameters tuning for RF and SVM models. Using the tuned parameters, the performance of the models was verified through 5-fold cross-validation. The RF model resulted in accuracy ranging from 71.9\% to 80.6\%. For the SVM model, adjustments in the $C$ and $\Gamma$ resulted in accuracy between 72.8\% and 84.7\% (Table \ref{tab:optimization_results}). The ROC plot (Fig. \ref{fig:auc}) exhibits above-chance classification ability for both models, with SVM showing superior performance. In our study, we employed the Optuna framework for hyperparameter optimization, utilizing TPE for adaptive Bayesian optimization. This method refines the search space dynamically based on feedback from ongoing trials, enhancing the precision in selecting optimal hyperparameters and reducing computational overhead. It allows for a nuanced exploration of hyperparameter effects, significantly improving model performance and generalization in complex EEG data analysis. 
This approach significantly differs from traditional methods like grid search, manual tuning, or nested cross-validation. The latter, used by Cooney et al. for EEG signal classification with CNNs, relies on extensive evaluations to optimize hyperparameters \cite{cooney2020evaluation}. Our approach introduces a novel aspect of EEG signal classification by leveraging an advanced, automated hyperparameter tuning process. This method enhances the precision in selecting optimal hyperparameters, reducing computational overhead, and leading to improved model generalization. This systematic and individualized tuning approach underscores the robustness and reliability of our findings, as each subject's model configuration is specifically tuned to derive the best possible outcomes from their data. 


\section{Conclusion}
In conclusion, our study developed a novel BCI platform for interpreting EEG signals in visual sustained attention tasks. We successfully extracted key temporal and spectral features. Our tuned SVM and RF models decoded participants' attentional state with notable accuracy of 80\% and 78\%, respectively. This research study is a foundation for creating a real-time closed-loop neurofeedback platform that is responsive to immediate neural pattern changes by employing high-speed open-source language programming and pipelines. 

\bibliographystyle{unsrt}
\bibliography{bib}

\appendices

\ifCLASSOPTIONcaptionsoff
  \newpage
\fi

\end{document}